\documentclass{revtex4-1}

\usepackage[latin1]{inputenc}
\usepackage{amsmath}

\usepackage{color}

\usepackage{graphicx}

\newcommand{\alert}[1]{{\color{red}#1}}

\begin{document}

\title{Minima of multi-Higgs potentials with triplets of $\Delta(3n^2)$ and $\Delta(6n^2)$}

\author{Ivo de Medeiros Varzielas}
\email{ivo.de@udo.edu}
\affiliation{
CFTP, Departamento de F\'{\i}sica, Instituto Superior T\'{e}cnico,\\
Universidade de Lisboa,
Avenida Rovisco Pais 1, 1049 Lisboa, Portugal
}

\begin{abstract}
I present a list of minima for 3 Higgs doublet models and 6 Higgs doublet models that are invariant under the discrete symmetries $\Delta(3n^2)$ and $\Delta(6n^2)$. I use the invariant approach formalism on the scalar sector and, with relevant Spontaneous CP-odd invariants, determine if the minima lead to Spontaneous CP violation. I identify in which cases the VEVs have calculable phases, finding several new cases of Spontaneous Geometrical CP Violation, including the first cases for 6 Higgs doublet models.
\end{abstract}

%\FullConference{Corfu Summer Institute 2017 "School and Workshops on Elementary Particle Physics and Gravity"\\
%         2-28 September 2017\\
%         Corfu, Greece}

\maketitle

\section{Introduction}

Multi-Higgs doublet models (MHDMs) are a well motivated Beyond Standard Model (BSM) scenario. One of its advantages over the Standard Model (SM) is the possibility for CP violation (CPV) arising from the scalar sector, explicitly or through vacuum expectation values (VEVs) that lead to Spontaneous CP violation (SCPV).

I will cover here the following:
\begin{enumerate}
\item Method to test for SCPV,
\item Method to find the VEVs,
\item For each potential, find the VEVs and if there is SCPV.
\end{enumerate}

This proceedings entry is based on several works \cite{Varzielas:2016zjc, deMedeirosVarzielas:2017glw, deMedeirosVarzielas:2017ote} studying the invariant approach to scalar potentials. The references therein are more complete. Some of methods used were developed significantly in \cite{Branco:2005em, Davidson:2005cw, Gunion:2005ja}. The basis-independent approach I describe here is extremely powerful. An approach using the special Higgs basis is discussed in \cite{Ogreid:2017alh}.

To start, we recast any potential of interest into a standard form. I consider here only renormalisable potentials that contain only quadratic and quartic terms (but generalising the potential to include e.g. cubic terms is fairly straightforward), and write:
\begin{equation}
V~=~ {\varphi^\ast}^a  \,Y_a^b\, \varphi_b 
+  {\varphi^\ast}^a{\varphi^\ast}^c  \,Z_{ac}^{bd}\, \varphi_b \varphi_d \,,
\end{equation}
where the $Z$ tensor by construction is symmetric under exchange of the two upper indices, and also under exchange of the two lower indices.
For the particular case of MHDMs, the $n$ Higgs doublets $H_{i  \alpha}=(h_{i,1},h_{i,2})$, where $\alpha=1,2$ denotes the $SU(2)_L$ index and $i$ goes from $1$ to $n$, can be arranged as follows
\begin{equation}
\varphi=(\varphi_1,\varphi_2, \ldots, \varphi_{2n-1},
\varphi_{2n})=(h_{1,1},h_{1,2},\ldots,h_{n,1},h_{n,2}) \,.
\end{equation}
In writing the potential in the standard form, invariance under symmetries does not need to be specified, as it is encoded in the $Y$ and $Z$ tensors.
For example, $SU(2)_L$ invariance restricts the quadratic terms to be $-m^2 \sum_{ \alpha}   h_{ \alpha}  h^{*\alpha}$:
$Y_1^1=Y_2^2 = -m^2$, $Y_1^2=Y_2^1 = 0$.

In terms of notation, I highlight the differences between complex conjugation, performing a basis change, or applying a general CP transformation.
Complex conjugation sends the scalars into their conjugates. I denote this with raising/lowering indices
\begin{align}
\varphi_a  \mapsto & (\varphi_a)^\ast \equiv  {\varphi^\ast}^a  \,,\\
{\varphi^\ast}^a  \mapsto &  ({\varphi^\ast}^a)^\ast \equiv {\varphi}_a \,.
\end{align}
The potential is real ($V=V^*$), implying that complex conjugation also raises and lowers the indices of the $Y$ and $Z$ tensors:
\begin{equation}
(Y_b^a)^\ast ~=~ Y_a^b \,,
\label{Y_real}
\end{equation}
\begin{equation}
(Z^{ac}_{bd})^\ast ~=~ Z_{ac}^{bd} \,.
\label{Z_real}
\end{equation}
I denote a basis change, acting on the scalars $\varphi_a$, with matrix $V$:
\begin{align}
\varphi_a \mapsto & V_a^{a'} \varphi_{a'} \\
{\varphi^\ast}^a \mapsto &  {\varphi^\ast}^{a'} {V^\dagger}_{a'}^a \,.
\end{align}
The notation with raised and lower indices is convenient, as the effect of the basis change on the tensors is simply applying the $V$ matrix for each index: 
\begin{align}
Y_a^b \mapsto & V_a^{a'} \, Y_{a'}^{b'} \, {V^\dagger}_{b'}^b \\
Z_{ac}^{bd} \mapsto & V_a^{a'}\, V_c^{c'} \, Z_{a'c'}^{b'd'} \,
{V^\dagger}_{b'}^b \,{V^\dagger}_{d'}^d \,. \label{eq:Zbasis}
\end{align}
A general CP transformation corresponds to sending each of the $\varphi_a$ scalars to conjugate scalars, but in general there is an associated unitary matrix that I denote as $X$:
\begin{align}
\varphi_a \mapsto & {\varphi^\ast}^{a'} X_{a'}^a \\
{\varphi^\ast}^a \mapsto & {X^\dagger}_a^{a'} \varphi_{a'} \,.
\end{align}
The use of raised and lowered indices is going to be quite convenient.

If $X$ is the unit matrix, then I have a trivial CP transformation which I refer to as $CP_0$.
As a non-trivial example of such a general CP transformation, take 
3 scalars $(\varphi_1, \varphi_2, \varphi_3)$. One may have a CP transformation with $X$ matrix:
\begin{align}
\label{eq:X_23}
X_{23} = \begin{pmatrix}
1 & 0 & 0 \\
0 & 0 & 1 \\
0 & 1 & 0 
\end{pmatrix} \,,
\end{align}
thus sending
$(\varphi_1, \varphi_2, \varphi_3) \mapsto X_{23} (\varphi^{*1}, \varphi^{*2}, \varphi^{*3}) = (\varphi^{*1}, \varphi^{*3}, \varphi^{*2})$.

Sometimes, general CP transformations are referred to as generalised CP transformations. The former nomenclature has the advantage that its spelling remains invariant under changes between U.K. and U.S. English.

\section{Spontaneous CP violation and new minima}

With the notation set, I now discuss specific potentials with 3 or 6 scalars which are invariant under certain discrete symmetries. For my purposes here it is sufficient to know the respective expressions (from which the corresponding $Y$ and $Z$ tensors can be obtained).

The simplest of the potentials that I will consider here is the potential invariant under $\Delta(6n^2)$, $n>3$, specified in the expression:
\begin{align}
\label{eq:V_0}
V_{\Delta(6n^2)} (\varphi) = V_0(\varphi) \equiv
 - ~m^2_{\varphi}\sum_i   \varphi_i \varphi^{*i}
+ r \left( \sum_i   \varphi_i \varphi^{*i}  \right)^2
+ s \sum_i ( \varphi_i \varphi^{*i})^2 \,.
\end{align}
This potential is invariant under both trivial CP transformation $CP_0$ and CP with $X_{23}$ ($i=1,2,3$).

Containing $V_0(\varphi)$, the potential invariant under $A_4$ is:
\begin{align}
V_{A_4} (\varphi) = V_0 (\varphi)+&c \left(
\varphi_1 \varphi_1 \varphi^{*3} \varphi^{*3} +  \varphi_2 \varphi_2 \varphi^{*1} \varphi^{*1} + \alert{\varphi_3 \varphi_3 \varphi^{*2} \varphi^{*2}} \right) \nonumber\\
+&c^\ast \left(
\varphi^{*1} \varphi^{*1} \varphi_3 \varphi_3 + \varphi^{*2} \varphi^{*2} \varphi_1 \varphi_1 + \alert{\varphi^{*3} \varphi^{*3} \varphi_2 \varphi_2}
 \right)
 \,.
\end{align}
$V_{A_4}(\varphi)$ is invariant under CP without any constraints, although it is not invariant under $CP_0$ in general - under $CP_0$, each of the terms multiplied by $c$ goes to a corresponding term multiplied by $c^*$, so this potential is only invariant under $CP_0$ if $c$ is real ($c=c^*$) . But for any $c$, the potential remains invariant under the general CP transformation associated with the $X_{23}$ matrix in eq.(\ref{eq:X_23}):
as 
$(\varphi_1, \varphi_2, \varphi_3) \mapsto X_{23} (\varphi^{*1}, \varphi^{*2}, \varphi^{*3}) = (\varphi^{*1}, \varphi^{*3}, \varphi^{*2})$,
of the three terms multiplied by $c$, the last (in red) remains invariant, and each of the other two terms goes into each other (the same applies for the terms multiplied by $c^*$). The potential is invariant under a CP transformation, and is therefore CP conserving.

Also containing $V_0(\varphi)$, the potential invariant under $\Delta(27)$ is:
\begin{align}
\label{eq:V_27}
V_{\Delta(27)} (\varphi)=
V_0(\varphi)
+&d \left(
\varphi_1 \varphi_1 \varphi^{*2} \varphi^{*3} + 
\text{cycl.} \right) \nonumber\\
+&d^* \left(
\varphi^{*1} \varphi^{*1} \varphi_{3} \varphi_{2} + 
\text{cycl.} \right)
\,,
\end{align}
where cycl. denotes the cyclic permutation of the 3 indices, e.g. $(\varphi_1  + \text{cycl.}) =(\varphi_1  + \varphi_2 + \varphi_3) $.
$V_{\Delta(27)}(\varphi)$ has CPV in general. It becomes CP conserving for special values of the parameters, e.g. it is 
invariant under both $CP_0$ and CP with $X_{23}$ if $d$ is real ($d=d^*$).

In order to study the CP properties of these and other potentials, it is necessary to consider that any general CP transformation that leaves the potential invariant guarantees the potential conserves CP. It is therefore convenient to employ methods based on basis invariants, $I$, that are independent of $V$ (basis change) and $X$ (general CP transformation). Such basis invariants can be built from $Y$ and $Z$ tensors (and also VEVs of the fields) by contracting all indices - as $V$ and $X$ are unitary matrices, the basis invariants will no longer depend on them as they cancel throughout.
For example, two basis invariants can be obtained by tracing the indices of a single $Z$ tensor
\begin{equation}
Z^{ab}_{ab}\text{ and }Z^{ab}_{ba} \,.
\end{equation}
They are of course invariant under basis change, e.g. we see from eq.(\ref{eq:Zbasis}):
\begin{equation}
Z^{ab}_{ab} \mapsto V_a^{a'}\, V_b^{b'} \, Z_{a'b'}^{a'b'} \,
{V^\dagger}_{a'}^a \,{V^\dagger}_{b'}^b = Z_{a'b'}^{a'b'} = Z^{ab}_{ab}\,.
\end{equation}
The invariance under basis change generalizes for any combinations, as long as all indices are contracted. Analogously, the dependence on $X$ when performing a general CP transformation is cancelled.

The number of basis invariants grows quickly as more tensors are combined, as all $n$ upper indices can be contracted with all $n$ lower indices in $n!$ ways (corresponding to each element of the permutation group $S_n$). There are already $4! = 24$ possibilities with two $Z$ tensors. But due to the properties of the $Z$ tensor and because many of the contractions can be expressed in terms of other invariants, of the 24 there are just two independent ones, e.g.:
\begin{equation}
Z^{ab}_{bd}Z^{cd}_{ac}\text{ and }Z^{ab}_{cd}Z^{cd}_{ab} \,.
\label{2Z_invariants}
\end{equation}
Many invariants are equivalent, or products of smaller invariants.

Further, many basis invariants are CP-even (all examples so far).
To study CP properties, I want to use \alert{CP-odd invariants} (CPIs)
\begin{equation}
 \mathcal{I}=I-I^{\ast} \,.
 \label{eq:CPI_I_c}
\end{equation}
Diagrams, and particularly a technique using coupling matrices introduced by \cite{Varzielas:2016zjc}, help find cases with \alert{$I \neq I^*$}. These CPIs can test if a potential has explicit CPV (without worrying about $X$s). Indeed, a non-vanishing CPI guarantees there is explicit CPV. Several CPIs are listed in \cite{Varzielas:2016zjc} (all CPIs up to 6 $Z$ tensors). Using the CPIs one can very elegantly study the explicit CPV properties of potentials, such as the multi-Higgs potentials with triplets of $\Delta(3n^2)$ and $\Delta(6n^2)$ analysed in \cite{Varzielas:2016zjc}.

In contrast, Spontaneous CPIs (SCPIs) are built from basis invariants $J$:
\begin{equation}
 \mathcal{J}=J-J^{\ast} \,,
\end{equation}
where $J$ involves the VEVs, $v_a \equiv \langle \varphi_a \rangle$.
These SCPIs can test if a VEV has SCPV, in a specific CP conserving potential. Indeed, a non-vanishing SCPI guarantees the corresponding VEV violates CP spontaneously, for the CP conserving potential being studied.

\begin{figure}
\begin{center}
\includegraphics[scale=0.20]{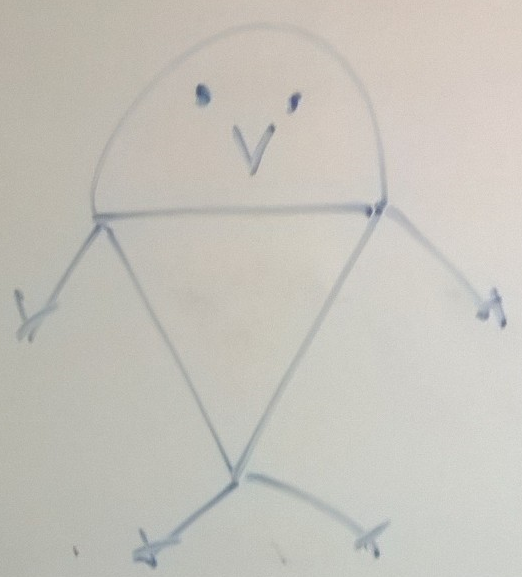}
\end{center}
\caption{The penguin diagram, as first depicted on a whiteboard. \label{fig:penguin}}
\end{figure}

A particularly useful SCPI is obtained from the basis invariant that I denote as $J^{(3,2)}$ (for 3 $Z$ tensors and 2 pairs of VEVs):
\begin{equation}
J^{(3,2)}\equiv Z^{a_1a_2}_{a_4a_5}Z^{a_3a_4}_{a_2a_6}Z^{a_5a_6}_{a_7a_8}v_{a_1}v_{a_3}v^{\ast a_7}v^{\ast a_8}=\vcenter{\hbox{\includegraphics[scale=0.2]{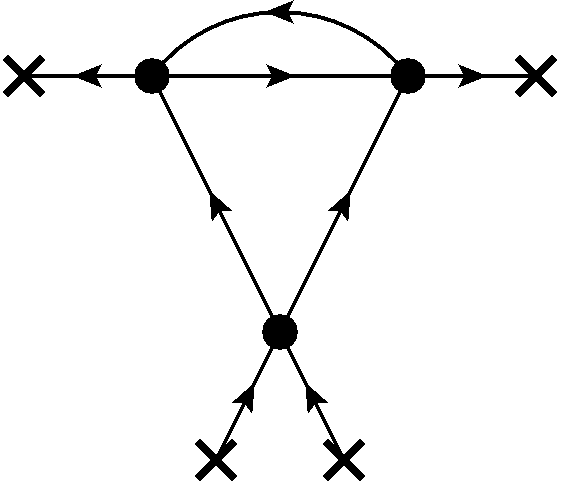}}} \,,
\label{J322}
\end{equation}
where in the associated diagram, which I sometimes refer to as the penguin diagram due to its resemblance to the flightless bird (see figure \ref{fig:penguin}), the dots denote a $Z$ tensor, the crosses denote a VEV, and the arrows indicate a contraction from an upper to a lower index. 
${J^{(3,2)}}^*$ is obtained by exchanging all upper and lower indices, and is associated with the diagram where all arrows are reversed, which is not equivalent (so the corresponding SCPI is indeed CP-odd).

As an example of the application of SCPIs, consider briefly the MHDM for 2 Higgs doublets $H_1$, $H_2$, when $CP_0$ (trivial CP) is imposed to make the potential CP conserving (SCPV makes sense if the potential is CP conserving):
\begin{equation}
\mathcal J^{(3,2)}_{CP_0} = F(a_1, a_2, b, b',c_1,c_2, d, \langle h_{i,j} \rangle) %9/16 (-2 a_2^2 c_1+2 a_1^2 c+2+(c_1-c_2) (c_1+c_2)^2+a_2 (c_1+c_2) (b+b'+2 d)+a_1 (2 a_2 (c_1-c_2)-(c_1+c_2) (b+b'+2 d))) 
\alert{[\langle h_{2,1} \rangle \langle h_{1,1} \rangle^*+ \langle h_{2,2} \rangle \langle h_{1,2} \rangle^* 
% - \langle h_{1,1} \rangle \langle h_{2,1} \rangle^* 
- h.c. ]}
%\langle h_{1,2} \rangle \langle h_{2,2} \rangle^*]
\,,
\end{equation}
where $F(a_1, a_2, b, b',c_1,c_2, d, \langle h_{i,j} \rangle)$ is a function of the coefficients $a_1, a_2, b, b',c_1,c_2, d$ of the quartic terms of the potential, shown in the appendix in eq.(\ref{2SU2}).
Most relevant is the functional dependence on the VEVs (in red), where the SCPI shows that SCPV depends on \alert{relative phase} between $\langle H_{1} \rangle$ and $\langle H_{2} \rangle$
(note $\langle h_{1,1} \rangle = \langle h_{2,1} \rangle =0$ for charge preserving VEVs).

I consider now $V_{\Delta(27)} (\varphi)$ again, in eq.(\ref{eq:V_27}). It contains $V_0(\varphi)$ (eq.(\ref{eq:V_0})), which has two quartics, with coefficients $r$ and $s$, and additionally the coefficient $d$. As I already mentioned, $V_{\Delta(27)}(\varphi)$ has CPV in general, and is invariant under $CP_0$ if $d$ is real.
When applying the SCPI to the potential (meaning one uses the corresponding $Z$ tensor), the general structure is thus:
\begin{eqnarray}
\mathcal{J}^{(3,2)} = A(d) Q(|v_i|) + B(d,s) R(v_i)\,,
\end{eqnarray}
where
$A(d) = 0$ for some CP symmmetris,
$R(v_i) \neq 0$ for VEVs that violate those; and
$B(d,s) =0 $ for some other CP syms.,
$Q(|v_i|) \neq 0$ for VEVs that violate those.
Even without specifying the functions, this structure is already automatically tracking that there are different CP symmetries making one of the two terms in the expression vanish. And further, that depending on the CP symmetry chosen, the term that does not vanish depends on the VEVs in such a way that, if the VEV conserves that CP symmetry, the remaining term also vanishes. This is not a coincidence, as a non-vanishing SCPI guarantees that there is SCPV.

In order to analyse this further, here is the expression
\begin{eqnarray}
  && \mathcal{J}^{(3,2)} = \frac{1}{4}\alert{(d^{\ast 3}-d^3)} Q(|v_i|)
\nonumber  \\&&+\frac{1}{2}(dd^{\ast 2}-2d^\ast s^2+d^{2}s)(v_2 v_3 v_1^{\ast 2}+v_1
  v_3 v_2^{\ast2}+v_1 v_2 v_3^{\ast2}) \nonumber\\
&&-\frac{1}{2}(d^2d^\ast-2ds^2+d^{\ast 2}s)(v_2^\ast v_3^\ast
 v_1^2+v_1^\ast v_3^\ast v_2^2+v_1^\ast v_2^\ast v_3^2)   \label{eq:J27ex2} \,,
\end{eqnarray}
where $Q(|v_i|)$ is a quartic function of the absolute values of the VEVs.
If I impose $CP_0$, I force $d$ to be real, \alert{$d=d^*$}.
Clearly this simplifies the expression:
\begin{align}
\label{eq:J32CP0D27}
&\mathcal{J}_{CP_0}^{(3,2)} = \frac{1}{2}(d^3 -2d s^2+d^{2}s) \nonumber \\
&\left[(v_2 v_3 v_1^{\ast 2}+v_1 v_3 v_2^{\ast2}+v_1 v_2 v_3^{\ast2})-(v_2^\ast v_3^\ast v_1^2+v_1^\ast v_3^\ast v_2^2+v_1^\ast v_2^\ast v_3^2) \right]
   \end{align}
and I now focus on the VEV dependence. We can consider the different candidate VEVs of this potential (which of these candidates are the actual VEVs depends on the region of the parameter space). Both these candidates include the geometrical phase, $\omega \equiv e^{i 2 \pi/3}$, $\omega$, whose value is independent of the quartic coefficients, and are: $\langle \varphi \rangle = (1,\omega, \omega^2)$ which conserves CP, and $\langle \varphi \rangle = (\omega,1,1)$.
That $\langle \varphi \rangle = (\omega,1,1)$. violates CP (with a geometrical phase) is certain, as it gives a non-zero value when inserted into $\mathcal{J}_{CP_0}^{(3,2)}$ in eq.(\ref{eq:J32CP0D27}). This is a case of spontaneous geometrical CP violation (SGCPV).
Conversely, it may appear strange that the other VEV candidate is CP conserving, as it has complex phases and we are considering $CP_0$, which is clearly violated by the complex phases. This illustrates the convenience of the invariant approach, as if there is no SCPV, all CPIs must vanish and the result is, of course, basis independent. As it turns out, this CP conserving VEV with complex phases is equivalent to the real VEV $(1,1,1)$, in the sense that they are related to by a $\Delta(27)$ transformation which leaves the potential invariant. The VEVs thus related have the same value for the potential and are said to belong to the same orbit \cite{deMedeirosVarzielas:2017glw, deMedeirosVarzielas:2017ote}. More generally, a general CP symmetry (related to $CP_0$ by a $\Delta(27)$ transformation) both leaves the potential invariant and is still preserved by the VEV $(1,\omega, \omega^2)$. One unbroken general CP symmetry is sufficient to guarantee CP is preserved in vacuum (even though $CP_0$ is broken).

%\subsection{New minima}

In order to test candidate VEVs for SCPV, I need to know them in the first place. For the case of $\Delta(27)$ discussed above, they were already known.
A method of decreasing symmetry was proposed in \cite{deMedeirosVarzielas:2017glw, deMedeirosVarzielas:2017ote}, and using it the previously known VEVs were confirmed and new minima were found, in particular for potentials with two triplets of $\Delta(3n^2)$ and $\Delta(6n^2)$ (with $n=2, 3$ and $n>3$).
For one triplet of $A_4, S_4, \Delta(27), \Delta(54)$,
this simple method reproduces results in \cite{Ivanov:2014doa}

To illustrate the idea, lets consider the simpler potential for $\Delta(6n^2)$ with $n>3$, in eq.(\ref{eq:V_0}), which had not been analysed previously.
The VEV candidates were found to be
\begin{equation}
v_1 (1,0,0),\quad v_2 (1,1,0),\quad v_3 (1,1,1) \,,
\end{equation}
as follows.

I consider first that part of the potential is invariant under the larger $U(3)$ symmetry.
\begin{equation}
 V_0=V_{U(3)}+\alert{V_{\Delta(6\infty^2)\times U(1)}} \,.
\end{equation}
Under just the $U(3)$ part of the potential, there would be a single orbit connected by the continuous symmetry. VEVs with the same magnitude but different directions have the same value of the potential.
The remaining terms
\begin{equation}
V_{\Delta(6\infty^2)\times U(1)} = s (\varphi_1 \varphi^{*1} \varphi_1 \varphi^{*1}+\varphi_2 \varphi^{*2} \varphi_2 \varphi^{*2}+\varphi_3 \varphi^{*3} \varphi_3 \varphi^{*3}) \,,
\end{equation}
do distinguish the directions and splits the single $U(3)$ orbit into 3 classes of directions:
\begin{equation}
 \{\begin{pmatrix}
    e^{i\eta}\\0\\0
   \end{pmatrix},\begin{pmatrix}
   0\\e^{i\eta}\\0
   \end{pmatrix},\begin{pmatrix}
    0\\0\\e^{i\eta}
   \end{pmatrix}
\},\{\begin{pmatrix}
      e^{i\eta}\\e^{i\zeta}\\0
     \end{pmatrix},\text{permut.}
\},\{\begin{pmatrix}
      e^{i\eta}\\e^{i\zeta}\\e^{i\theta}
     \end{pmatrix},\text{permut.}
\} \,,
\end{equation}
where permut. signifies the possible permutations. Note there are relations between the magnitudes, but not the phases.

\begin{figure}
\begin{center}
      \resizebox{9 cm}{!}{
        \includegraphics{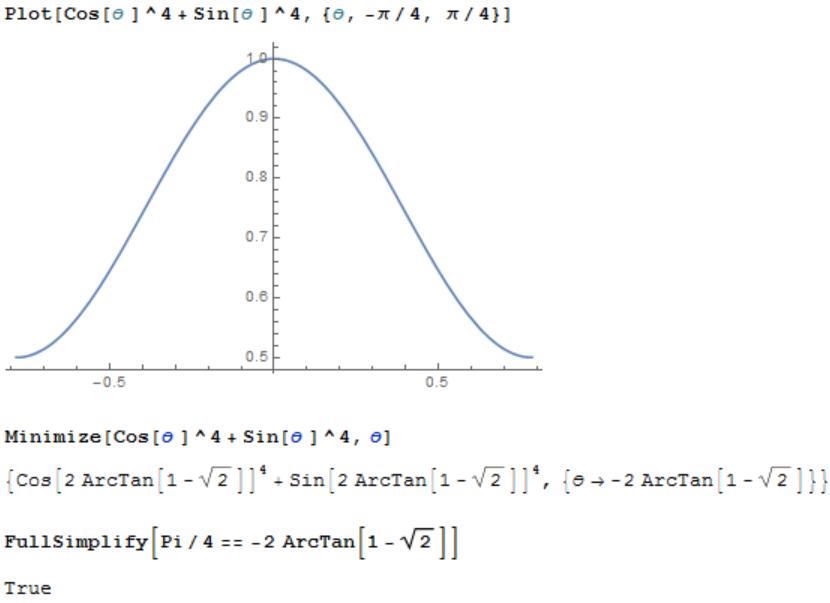}
      }
\end{center}
\caption{Calculation of extrema for 2 generation example. \label{fig:plot}}
    \end{figure}

To clarify this it is simpler to consider briefly a case where there are only 2 generations, where what I would look as VEV candidates would be the extrema of $s (|\varphi_{1}|^{4}  + |\varphi_{2}|^{4})$ for fixed magnitude $v$. Parametrising the direction with an angle $\theta$, I find the candidates $v(\cos \theta, \sin \theta)$ to be, depending on the sign of the $s$ coefficient:\\
positive $s$, VEV $\propto (1,1)/\sqrt{2} \rightarrow V \sim +2 v^{4}/4$,\\
negative $s$: VEV $\propto (1,0)~\text{or}~(0,1) \rightarrow V \sim -v^{4}$.\\
This can be visualized in figure \ref{fig:plot}.

Using this method, the orbits are progressively split, by starting with the larger symmetry and adding terms that are only invariant under the smaller symmetry.
For two triplets of $\Delta(3n^2)$ (with $n>3$), the method reveals some candidate VEVs like
\begin{align*}
 (e^{i\eta},0,0),(e^{i\eta'},0,0)&\rightarrow (1,0,0),(1,0,0)\\
 (e^{i\eta},e^{i\zeta},e^{i\theta}),(e^{i\eta'},e^{i\zeta'},e^{i\theta'})&\rightarrow (1,1,1),(1,e^{i\alert{\zeta'}},e^{i\alert{\theta'}}).
\end{align*}
For the last case, the phases (in red) are physical. Following the method, I minimize the phase-dependent part of potential to fix the phases.
That reveals new VEVs with geometrical phases
\begin{equation*}
 (1,1,1),(1,\omega,\omega^2)\text{ and }(1,1,1),(1,\omega^2,\omega).
\end{equation*}
Do these VEVs have SCPV?

I use the same SCPI, now applied to the respective potential (the potential is shown in the appendix in eqs.(\ref{eq:potV1}, \ref{V3n2PP})):
\begin{align}
 \mathcal J^{(3,2)}=
&-\frac{1}{16} \tilde{s}_2 [ \tilde{r}_2  (-4 s-4 s'+2 \tilde{s}_1-\tilde{s}_2+3 \tilde{r}_2) - \tilde{s}_3^2 ] W_{CP_0} \nonumber \\
&-\frac{1}{8} i \alert{\tilde{s}_3} \left[\tilde{s}_3^2-3 \tilde{r}_2^2\right] W_{CP_{23}} \nonumber \\
&-\frac{1}{16} i \tilde{s}_2 \alert{\tilde{s}_3} [-4s - 4 s' + 2 \tilde{s}_1 -\tilde{s}_2] (...) \,,
\end{align}
where
\begin{align*}
W_{CP_0} \equiv [ ( v_1 {v'}_1^\ast {v'}_2 v_2^\ast + v_2 {v'}_2^\ast {v'}_3  v_3^\ast + v_3 {v'}_3^\ast {v'}_1 v_1^\ast) - h.c.]\,,
\end{align*}
and $CP_0$ forces $\alert{\tilde{s}_3 = 0}$ (this is the coefficient of one of the quartics terms, the only one with a complex phase present, see eq.(\ref{eq:potV1})).

Given the dependence:
\begin{align}
W_{CP_0} \equiv [ ( v_1 {v'}_1^\ast {v'}_2 v_2^\ast + v_2 {v'}_2^\ast {v'}_3  v_3^\ast + v_3 {v'}_3^\ast {v'}_1 v_1^\ast) - h.c.]\,,
\end{align}
we see that
\begin{align}
W_{CP_0}^{(3,2)}[v(1,1,1),v'(1,\omega,\omega^2)] = 
3 (\omega - \omega^2) v^2 {v'}^2 \neq 0 \,.
\end{align}
A non-zero SCPI means there is SCPV. This means \alert{there are new cases of SGCPV in $\Delta(3n^2) \times CP_0$, for six scalars. The analogous version for Higgs doublets reveals the same for the corresponding 6HDM.}
A fuller analysis including tables of highlighting such cases is present in \cite{deMedeirosVarzielas:2017ote}.

\section{Conclusion}

In conclusion, I presented work where the formalism for CPIs and SCPIs, and for finding candidate minima was developed.
Using these methods for explicit and spontaneous CP violation, one can analyse any potential when brought to standard form.
The CP properties of 3HDM and 6HDM symmetric under $\Delta(3n^2)$ and $\Delta(6n^2)$ groups were verified, for new minima that were found, which include new cases with SGCPV.

\appendix

\section{Potentials}

\begin{eqnarray}
V (H_{1},H_{2})&=&m_{1}^2 \ H _{1}^{\dagger }H _{1}+ m_{12}^2\ e^{i\theta_0 }\ H
_{1}^{\dagger }H _{2}+ m_{12}^2 \ e^{-i\theta_0 }\ \ H _{2}^{\dagger }H
_{1}+m_{2}^2\ H _{2}^{\dagger }H _{2}+ \nonumber \\[2mm]
&&+a_{1}\ \left( H _{1}^{\dagger }H _{1}\right) ^{2}+a_{2}\ \left( H
_{2}^{\dagger }H _{2}\right) ^{2}\nonumber \\[2mm]
&&+b\ \left( H _{1}^{\dagger }H
_{1}\right) \left( H _{2}^{\dagger }H _{2}\right) +b^{\prime }\ \left(
H _{1}^{\dagger }H _{2}\right) \left( H _{2}^{\dagger }H
_{1}\right) + \nonumber \\[2mm]
&&+c_{1}\ e^{i\theta _{1}}\ \left( H _{1}^{\dagger }H _{1}\right) \left(
H _{2}^{\dagger }H _{1}\right) +c_{1}\ e^{-i\theta _{1}}\ \left( H
_{1}^{\dagger }H _{1}\right) \left( H _{1}^{\dagger }H _{2}\right) +
\nonumber \\[2mm] 
&&+c_{2}\ e^{i\theta _{2}}\ \left( H _{2}^{\dagger }H _{2}\right) \left(
H _{2}^{\dagger }H _{1}\right) +c_{2}\ e^{-i\theta _{2}}\ \left( H
_{2}^{\dagger }H _{2}\right) \left( H _{1}^{\dagger }H _{2}\right) +
\nonumber \\[2mm]
&&+d\ e^{i\theta_3 }\ \left( H _{1}^{\dagger }H _{2}\right) ^{2}+d\
e^{-i\theta_3 }\ \left( H _{2}^{\dagger }H _{1}\right) ^{2}.
\label{2SU2}
\end{eqnarray}

\begin{eqnarray}
V_1 (\varphi,\varphi') &=&
+~ \tilde r_1 \left( \sum_i \varphi_i \varphi^{*i} \right)
\left( \sum_j \varphi'_j \varphi'^{*j} \right) 
+ \tilde r_2\left( \sum_i \varphi_i \varphi'^{*i} \right)
\left( \sum_j \varphi'_j \varphi^{*j} \right) \notag \\[2mm]
&& +~ \tilde s_1\sum_i \left(\varphi_i \varphi^{*i} \varphi'_i \varphi'^{*i}
\right) \notag \\[2mm]
&& +~ \tilde s_2 \left(
\varphi_1 \varphi^{*1} \varphi'_2 \varphi'^{*2} + 
\varphi_2 \varphi^{*2} \varphi'_3 \varphi'^{*3} + 
\varphi_3 \varphi^{*3} \varphi'_1 \varphi'^{*1} 
\right)  \notag \\[2mm]
&& +~ i \, \tilde s_3 
\Big[
(\varphi_1 \varphi'^{*1} \varphi'_2 \varphi^{*2} + \text{cycl.}
) 
- 
( \varphi^{*1}\varphi'_1  \varphi'^{*2} \varphi_2 +\text{cycl.}
)
\Big].
\label{eq:potV1}
\end{eqnarray}

\begin{eqnarray}
V_{\Delta(3n^2)} (\varphi,\varphi') &=&
V_0 (\varphi) + V_0'(\varphi') + V_1 (\varphi, \varphi'),
\label{V3n2PP}
\end{eqnarray}


\begin{thebibliography}{99}

%\cite{Varzielas:2016zjc}
\bibitem{Varzielas:2016zjc}
  I.~de Medeiros Varzielas, S.~F.~King, C.~Luhn and T.~Neder,
  %``CP-odd invariants for multi-Higgs models: applications with discrete symmetry,''
  Phys.\ Rev.\ D {\bf 94} (2016) no.5,  056007
  doi:10.1103/PhysRevD.94.056007
  [arXiv:1603.06942 [hep-ph]].
  %%CITATION = doi:10.1103/PhysRevD.94.056007;%%
  %12 citations counted in INSPIRE as of 13 Apr 2018


%\cite{deMedeirosVarzielas:2017glw}
\bibitem{deMedeirosVarzielas:2017glw}
  I.~de Medeiros Varzielas, S.~F.~King, C.~Luhn and T.~Neder,
  %``Minima of multi-Higgs potentials with triplets of $\Delta(3n^2)$ and $\Delta(6n^2)$,''
  Phys.\ Lett.\ B {\bf 775} (2017) 303
  doi:10.1016/j.physletb.2017.11.005
  [arXiv:1704.06322 [hep-ph]].
  %%CITATION = doi:10.1016/j.physletb.2017.11.005;%%
  %4 citations counted in INSPIRE as of 13 Apr 2018


%\cite{deMedeirosVarzielas:2017ote}
\bibitem{deMedeirosVarzielas:2017ote}
  I.~de Medeiros Varzielas, S.~F.~King, C.~Luhn and T.~Neder,
  %``Spontaneous CP violation in multi-Higgs potentials with triplets of $\Delta(3n^2)$ and $\Delta(6n^2)$,''
  JHEP {\bf 1711} (2017) 136
  doi:10.1007/JHEP11(2017)136
  [arXiv:1706.07606 [hep-ph]].
  %%CITATION = doi:10.1007/JHEP11(2017)136;%%
  %4 citations counted in INSPIRE as of 13 Apr 2018


%\cite{Branco:2005em}
\bibitem{Branco:2005em}
  G.~C.~Branco, M.~N.~Rebelo and J.~I.~Silva-Marcos,
  %``CP-odd invariants in models with several Higgs doublets,''
  Phys.\ Lett.\ B {\bf 614} (2005) 187
  doi:10.1016/j.physletb.2005.03.075
  [hep-ph/0502118].
  %%CITATION = doi:10.1016/j.physletb.2005.03.075;%%
  %79 citations counted in INSPIRE as of 13 Apr 2018


%\cite{Davidson:2005cw}
\bibitem{Davidson:2005cw}
  S.~Davidson and H.~E.~Haber,
  %``Basis-independent methods for the two-Higgs-doublet model,''
  Phys.\ Rev.\ D {\bf 72} (2005) 035004
   Erratum: [Phys.\ Rev.\ D {\bf 72} (2005) 099902]
  doi:10.1103/PhysRevD.72.099902, 10.1103/PhysRevD.72.035004
  [hep-ph/0504050].
  %%CITATION = doi:10.1103/PhysRevD.72.099902, 10.1103/PhysRevD.72.035004;%%
  %249 citations counted in INSPIRE as of 13 Apr 2018


%\cite{Gunion:2005ja}
\bibitem{Gunion:2005ja}
  J.~F.~Gunion and H.~E.~Haber,
  %``Conditions for CP-violation in the general two-Higgs-doublet model,''
  Phys.\ Rev.\ D {\bf 72} (2005) 095002
  doi:10.1103/PhysRevD.72.095002
  [hep-ph/0506227].
  %%CITATION = doi:10.1103/PhysRevD.72.095002;%%
  %144 citations counted in INSPIRE as of 13 Apr 2018


%\cite{Ogreid:2017alh}
\bibitem{Ogreid:2017alh}
  O.~M.~Ogreid, P.~Osland and M.~N.~Rebelo,
  %``A Simple Method to detect spontaneous CP Violation in multi-Higgs models,''
  JHEP {\bf 1708} (2017) 005
  doi:10.1007/JHEP08(2017)005
  [arXiv:1701.04768 [hep-ph]].
  %%CITATION = doi:10.1007/JHEP08(2017)005;%%
  %2 citations counted in INSPIRE as of 13 Apr 2018


%\cite{Ivanov:2014doa}
\bibitem{Ivanov:2014doa}
  I.~P.~Ivanov and C.~C.~Nishi,
  %``Symmetry breaking patterns in 3HDM,''
  JHEP {\bf 1501} (2015) 021
  doi:10.1007/JHEP01(2015)021
  [arXiv:1410.6139 [hep-ph]].
  %%CITATION = doi:10.1007/JHEP01(2015)021;%%
  %31 citations counted in INSPIRE as of 13 Apr 2018

\end{thebibliography}
\end{document}